# Pancharatnam-Berry Phase Induced Spin-Selective Transmission in Herringbone Dielectric Metamaterials

*Mitchell Kenney[1, 4\*], Shaoxian Li[2\*], Xueqian Zhang[2], Xiaoqiang Su[2], Teun-Teun Kim[1], Dongyang Wang[1, 2], Dongmin Wu[3], Chunmei Ouyang[2†], Jiaguang Han[2], Weili Zhang[2], Hongbo Sun[5‡] and Shuang Zhang[1, 2§]*

1. School of Physics & Astronomy, University of Birmingham, B15 2TT, UK

2. Center for Terahertz Waves and College of Precision Instrument and Optoelectronics Engineering, Tianjin University, Tianjin 300072, China

3. Key Laboratory of Nanodevices, Suzhou Institute of Nano-tech and Nano-bionics, Chinese Academy of Sciences, 398 Ruoshui Road, Suzhou, Jiangsu 215123, China

4 School of Engineering, Rankine Building, University of Glasgow, G12 8LT, UK

5. State Key Laboratory on Integrated Optoelectronics, College of Electronic Science and Engineering, Jilin University, Changchun 130012, China

**Abstract**

**Manipulating the polarisation of light is crucial for sensing and imaging applications. One such aspect in particular is selective transmission of one circular polarisation (spin) when light is transmitted through a medium or a device. However, most present methods of achieving this have relatively low efficiency and selectivity, whilst high selectivity examples rely on lossy and complex three-dimensional helical or multilayer structures. Here, we propose a dielectric metamaterial approach for achieving spin-selective transmission of electromagnetic waves, utilizing spin-controlled constructive or destructive interference between two Pancharatnam-Berry (PB) phases in conjunction with propagative dynamic phase. The dielectric metamaterial, consisting of monolithic silicon herringbone structures, exhibits a broadband operation in the terahertz regime**

---

[\*] These authors contributed equally to this work.
[†] cmouyang@tju.edu.cn
[‡] hbsun@jlu.edu.cn
[§] s.zhang@bham.ac.uk

**whilst obtaining a spin-selective efficiency upwards of 60%. Such a device is robust and is not easily degraded by errors in fabrication.**

A key goal in the development of light-based systems is being able to accurately control the polarisation of light. This is typically achieved with conventional polarisers and wave plates; however, these can be very bulky and not scalable to subwavelength sized devices. In the past decade or so, metamaterials have shown great promise in the controllability of light, to realise phenomena such as negative refraction [1–4], zero refractive index [5–7], invisibility cloaking [8–10], and sub-diffraction superlensing [11–13]. Even though metamaterials have been essential in developing new devices and a better insight into various fundamental physical laws, they still have not been adapted for real world applications. This is mainly due to the difficulties in fabrication and inherently low efficiency, which is a critical necessity in the ever growing requirement for lower energy consumption of devices. These issues are not trivial to overcome, with most metamaterials typically being composed of lossy metals, especially in the visible spectrum, and involving precise and time-consuming fabrication alignment processes, which is not an easy task even for the most state-of-the-art fabrication tools.

Recently, chiral metamaterials in the absence of mirror symmetry have been developed to invoke a chiral response, namely a contrast between the transmitted opposite spins of light. Most of these chiral metamaterials consist of 3D continuous helical structures [14,15] or stacked [16,17] metallic structures with twisted orientations. These 3D chiral structures (such as a helix) appear identical when viewed from both forward and backward directions and thus have equal responses for light. This in turn allows preferential transmission of one cross-polarisation whilst prohibiting or reflecting the opposite one. Conversely, as detailed in the original work in [18] an anisotropic lossy planar chiral "fish-scale" structure was investigated and shown to exhibit Asymmetric Transmission (AT) (*alternatively known as Circular Conversion Dichroism (CCD)*) for circularly polarised (CP) light in the microwave region, which was explained by the 'twist' vector **W** following the well-known 'cork-screw' law. Such a design appears reversed when viewed from opposite sides and so exhibits a directional response for the handednesses of CP light. This was then scaled down to work in the visible spectrum [19]. Similar works on using planar chiral metasurfaces were also carried out in the terahertz regime [20], and for investigating broadband capabilities in the infrared (IR) [21]. However, the chiral responses reported are usually very small for planar chiral metasurfaces. Recent methods have aimed to improve on this by utilizing aforementioned layered metasurfaces [22,23]. However, these devices have very complex designs, involving time consuming optimizations of layer-to-layer distance and impedance matching, as well as difficult fabrication processes. In addition, these devices are still composed of lossy metals, and thus absorption losses are unavoidable (where in [23] losses are 37%).

Here, we present a new means of achieving a strong chiral response through a spin-selective interference between light of different PB phases inside a birefringent metamaterial grating. The binary grating consists of a periodic arrangement of two uniaxial birefringent materials with the same anisotropy but an orientation angle difference θ = π/4 (each one forms a π/8 angle with respect to the symmetry line). The thickness of each of the birefringent materials satisfies the half wave plate condition such that a circularly polarised beam is completely converted to its opposite spin upon transmission. This spin flipping introduces a PB phase [24] such that the phase change of light introduced by an orientation /2, with the + sign corresponding to RCP incidence/LCP transmission and the − sign corresponding to LCP incidence/RCP transmission. In addition, one structure is elevated with a certain thickness relative to the other such that a π/2 difference in the dynamic phase between the two structures is introduced due to the different refractive indices between the substrate and the incident medium (air). Hence the overall phase difference between the two structures experienced by different handedness of light is then

$$\Phi = \varphi_{Dynamic} \pm \varphi_{Geometric} = \frac{\pi}{2} \pm \frac{\pi}{2} \qquad (1)$$

which yields a total phase change of $\Phi = 0$ for LCP → RCP, and $\Phi = \pi$ for RCP → LCP. Therefore, RCP incident light undergoes destructive interference, *resulting in no LCP transmission and being completely reflected*, whilst LCP incidence has no overall phase change, corresponding to constructive interference and anti-reflection behaviour, *therefore allowing full transmission of converted RCP light*. A diagram of the functional device is shown in **Figure 1c**. This novel approach is a simple yet effective method of achieving the aforementioned chiral response of CP light to yield a similar result as metallic chiral metasurfaces, yet only requiring dielectric materials.

It is well studied that subwavelength gratings can be utilized to exhibit birefringence [25], whereby the incident light experiences both fast and slow axes dependent upon its polarisation states (TE and TM). This introduces a form-birefringence to alter the polarisation state of the transmitted light, due to the grating refractive indices $n_{TM}$ and $n_{TE}$; the grating is then equivalent to a conventional birefringent crystal such as calcite, as a result of the effective medium approximation. By utilizing birefringence, the resulting analogy of the spin-selective metamaterial is given in **Figure 1a**, showing two identical birefringent crystals which are rotated such that the angle between their fast axes is $2\alpha = 45°$. A simplified schematic of the functionality of such a device is given in **Figure 1b**, where RCP incident light undergoes a phase change of $\pi$ and therefore is not transmitted (only reflected), whilst an incident LCP wave is flipped and undergoes a phase change of $0\pi$ (or $2\pi$) and therefore transmission is maximised. Such a functionality is then utilised via subwavelength (form-birefringent) gratings using silicon. This design has key advantages over the majority of previous methods used to achieve similar responses — the device is entirely made only of dielectric material, therefore Ohmic losses are negligible; fabrication is easily achieved by conventional plasma etching and photolithography; a simple and robust functionality is used, based on standard form-birefringent gratings in conjunction with geometric phase.

We investigated this effect at a frequency of 1 THz, within the so called "terahertz gap", as devices are of particular technological importance in this regime. To realise this, we chose to use silicon (Intrinsic, $\Omega$ = 10k Ohm, n = 3.418) which is transparent, has very low loss at THz frequencies, and is well studied for use in fabrication. The functionality of a subwavelength device is dependent upon the period of the features being smaller than the wavelength of interest. For subwavelength gratings, the periodicity must satisfy the equation:

$$\Lambda \leq \lambda/n_{II} \qquad (2)$$

where $n_{II}$ is the refractive index of the substrate material (and assuming that air is the background media $n_I = 1$), $\Lambda$ is the period of the gratings, and $\lambda$ is simply the free space wavelength of interest. According to equation (2), for a wavelength of 300 μm (corresponding to 1 THz) and the refractive index of silicon being $n_{Si} = 3.418$, the periodicity of the gratings must be $\Lambda \leq 87.8$ μm – to this end, we chose a periodicity of $\Lambda = 86$ μm. Providing that these gratings are subwavelength, the equation describing the depth and duty-cycle dependency of the phase accumulated by the light is given as:

$$\Delta\Phi_{TE-TM}(\lambda) = \left(\frac{2\pi h}{\lambda}\right)\Delta n_{form}(\lambda) \qquad (3)$$

where $h$ is the height/depth of the gratings (as shown in Fig.3), $\lambda$ is the free-space wavelength of the incident light, and $\Delta n_{form} = n_{TE}(\lambda) - n_{TM}(\lambda)$ is the difference between the refractive indices for light parallel (TE) or perpendicular (TM) to the gratings, respectively, which are given by:

$$n_{TE} = (Fn_I^2 + (1-F)n_{II}^2)^{1/2} \tag{4}$$

$$n_{TM} = (Fn_I^{-2} + (1-F)n_{II}^{-2})^{-1/2} \tag{5}$$

and $F$ is simply the duty cycle of the gratings. Using the previously given refractive indices of air and silicon as $n_I = 1$ and $n_{II} = 3.418$, respectively, we have that $n_{TE} = 2.52$ and $n_{TM} = 1.36$, respectively. We then have that $\Delta n_{form}(\lambda) = 1.16$, such that when Equation (3) is rearranged we obtain a grating depth of $h = 129\ \mu m$ (at a wavelength of 300 $\mu m$ and using a phase difference of $\Delta\Phi = \pi$) for the SWGs to function as half-wave plates. The schematic diagram of our device is shown in **Figure 1c**. We incorporate the Geometric Phase, to supply the necessary handedness dependent $\pi/2$ phase shift, by having an angle of 45° set between the TM axes of two SWGs. The additional $\pi/2$ dynamic phase is introduced by an extra 'step' of silicon beneath one such grating, and is calculated using

$$\Delta\varphi_{Dynamic} = \Delta n_{Si-Air}(2\pi d/\lambda) \tag{6}$$

where $d$ is required to be 31 μm ($\Delta n_{Si-Air} = n_{Si} - n_{Air} = 2.418$). Incorporating all of these calculated dimensions together gives us our Monolithic silicon Herringbone patterned device.

To further support our theoretical predictions, a simplified analytical model based on Fresnel's equations for transmittance was employed. The system was considered to have three-layers – with layer 1 being air, layer 2 being an SWG, and layer 3 being bulk silicon. From this, we used the Fresnel equation for transmittance:

$$t_i = \frac{t_{12_i} t_{23_i} e^{-i\phi}}{1 + r_{12_i} r_{23_i} e^{-2i\phi}} \tag{7}$$

where $i$ corresponds to $x$ or $y$ unit vectors, $t_{12_i} = 2n_1/(n_1 + n_{2i})$, $t_{23_i} = 2n_{2i}/(n_{2i} + n_3)$, $r_{12_i} = (n_1 - n_{2i})/(n_1 + n_{2i})$, $r_{23_i} = (n_{2i} - n_3)/(n_{2i} + n_3)$, and $\phi_i = \frac{2\pi d}{\lambda} n_{2i}$, where $d$ is the thickness of the SWG (layer 2). Once the circular Jones components for a single SWG are obtained, we then introduce the geometric and dynamic phase accumulations through a second identical SWG. The total transmitted intensity of the two SWG's combined is then given as:

$$T = n_3 \left| \frac{1}{2} t \left(1 + e^{i(\phi_{Dyn} + \phi_{Geom})}\right) \right|^2 \tag{8}$$

with *t* corresponding to any of the circular Jones matrix components for the SWG ($t_{RR}$, $t_{LL}$, $t_{LR}$, $t_{RL}$), $\phi_{Dyn}$ is simply the dynamic phase from Equation 6, and $\phi_{Geom}$ is the geometric phase equal to $\pm \pi / 2$ (depending on the cross-polarisation component) arising from the angular disparity of the two SWG's ($\phi_{Geom} = 0$ for both $t_{RR}$ and $t_{LL}$). The calculated frequency dependent response of the transmitted intensities are shown in **Figure 2a**. The values of *d* and *h,* as shown in Figure 1c, were set to 31 μm and 129 μm, respectively, as required for optimal functionality. A very clear difference between both $T_{RL}$ and $T_{LR}$ components can be seen, where the intensity of $T_{RL}$ exceeds 85% transmittance whilst $T_{LR}$ is negligible and has a value of zero transmittance at 1.0THz, as required. There is a slight difference between the frequencies at which the destructive ($T_{LR}$, $\Phi = \pi$) and unaffected ($T_{RL}$, $\Phi = 0$) beams occur, with $T_{LR} = 0$ occurring at the expected frequency of 1.0THz whilst the $T_{RL}$ maxima occurs at ~1.1 THz. This discrepancy can be attributed to the half-wave plates being approximated as SWG's form-birefringent from the First Order effective medium theory, rather than using a higher order formulation [26]. Additionally, the intuitive single-pass formulation does not take into consideration the interfacial aspects of the complete structure; hence, Fabry-Pérot resonance effects resulting from the reflectance terms in the denominator of Equation 7 lead to the analytical transmittances differing from the simple phase-only predictions of Equation 1.

In order to reinforce the theoretical reasoning and functionality of our device, 3D Finite-Difference Time Domain (FDTD) simulations were carried out using the commercially available *CST Microwave Studio* software package. The results of the circular transmission components are shown in **Figure 2b**. As can be seen, the results show a very good correspondence to those for the analytical model, especially apparent for $T_{RL}$ exceeding a transmittance of 85% at 1.05 THz, and also show a clear difference between the cross-polarisation components of $T_{RL}$ and $T_{LR}$. One discrepancy that is worth noting is that the device can no longer be considered as subwavelength for frequencies much larger than the operational frequency of ~1 THz, and would result in diffraction occurring causing spurious interference effects.

To experimentally verify our theoretical reasoning, the structure shown in **Figure 1c** was fabricated by conventional photolithography and plasma etching using a two-step pattern process (details provided in the Supplementary Information). The complete fabricated device is shown in **Figure 3**, imaged using a Scanning Electron Microscope (SEM). The measured geometry of the fabricated structure slightly differs to the ideal geometry; it is visible from the Inset of **Figure 3** that the top 'step' is larger than the bottom 'step', due to the decreased etch rate at the bottom of the trenches. As such, the top step is as designed at 31 μm whilst the bottom step is only 20 μm, which is a 35% difference in size. To establish the effect this has on the performance and functionality, simulations were carried out to reflect these different dimensions as shown in **Figure 4a**. It is clear that by comparing Fig.4a and Fig.2b we see very little difference in the transmission coefficient responses, with especially little difference to the curve for $T_{RL}$. Therefore, we can assume that the device will be highly robust to fabrication variations and even a 35% change in the desired step height does not significantly affect the physical response.

To characterise and obtain the transmission data for our device, a fiber-based Terahertz Time-Domain Spectroscopy (THz-TDS) system is used to measure linear Jones matrix components ($T_{xx}, T_{yy}, T_{xy}, T_{yx}$) of the herringbone structure at normal incidence for a frequency range of 0.2 - 2.0 THz, which can be converted into the transmission matrix in the circularly polarised basis. The results for the experimental data characterisation are shown in **Figure 4b**. It is clear that there is indeed an asymmetry between the $T_{RL}$ (red) and $T_{LR}$ (blue) components as expected. The frequency of highest conversion occurs at 1.025 THz, which is very close to the design frequency of 1 THz. Here, the value of $T_{RL}$ is 0.62, which indicates that 62% of LCP light incident onto the structure is converted into the opposite handedness of RCP. Conversely, $T_{LR}$ gives a value of only 0.13. The extinction ratio between $T_{RL}$ and $T_{LR}$ is then approximately 5:1, which is very close to the performance of that for the simulation. Although the performance of 62% for $T_{RL}$ is not quite the 85% as was achieved in the simulation, this result is still remarkable given that the structure is a single layer and purely dielectric. The discrepancy can be attributed to the non-negligible material losses and scattering which may occur, as well as fabrication errors. At this central frequency of 1.025 THz, a FWHM of 0.72 THz is achieved which is again very broadband, although not as much as that for the simulation. Examining the total energy of the device for both RCP and LCP incident handednesses, we get values of $T_R = 0.28$ and $T_L = 0.84$. Such a result is significant in the sense that this device can be used to distinguish the circular handedness of incident light to a very high efficiency from either transmission or reflection modes.

In conclusion, we have demonstrated and fabricated a functional monolithic dielectric device to achieve a strong asymmetry between the orthogonal circular polarisations of transmitted light, which provides a very high broadband capability and transmittance of one cross-polarisation whilst prohibiting the opposite one. Impressively, the herringbone metasurface not only provides an experimental cross-polarisation transmittance of 0.62 for $T_{RL}$, but also has a greater transmittance than pure silicon alone or any other similar works carried out using planar metasurfaces to achieve a chiral response. Due to the lack of metallic structures, losses are negligible and the application of subwavelength gratings in conjunction with a geometric phase provides a robust and novel means of achieving such a functionality, which may provide a preferable route for optical computing or image processing where the demand on high efficiency is crucial.

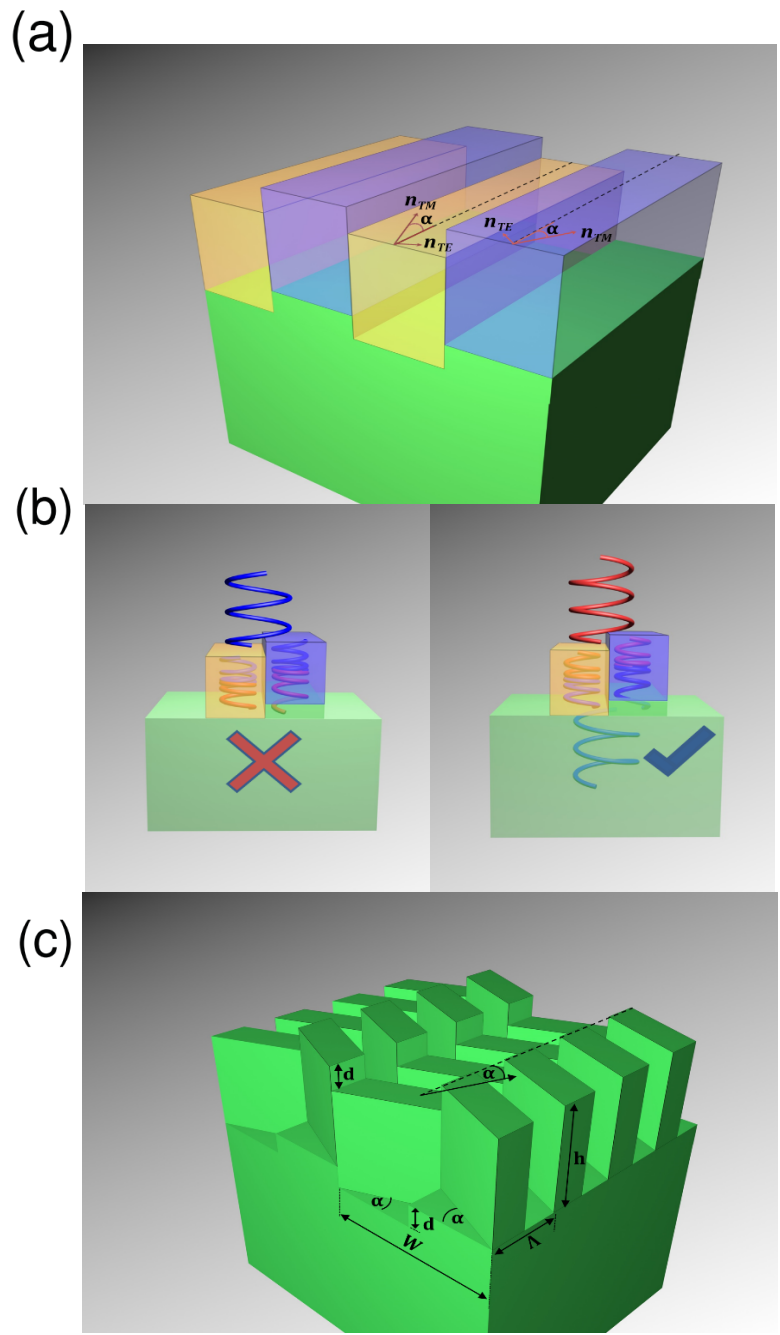

Figure 1. (a) Analogous representation of the herringbone device functionality, composed of homogeneous birefringent crystals with an angle of 45° between their fast axes (TM). (b) Schematic of the device functionality for RCP (blue) and LCP (red) is obtained, and limits transmission (only reflection), whilst for LCP incidence the ) and the transmitted light is flipped. (c) The monolithic silicon herringbone device and corresponding dimension parameters, where $\Lambda = 86\mu m$, $d = 31\mu m$, $W = 208\mu m$, $h = 129\mu m$, and $\alpha = 22.5°$.

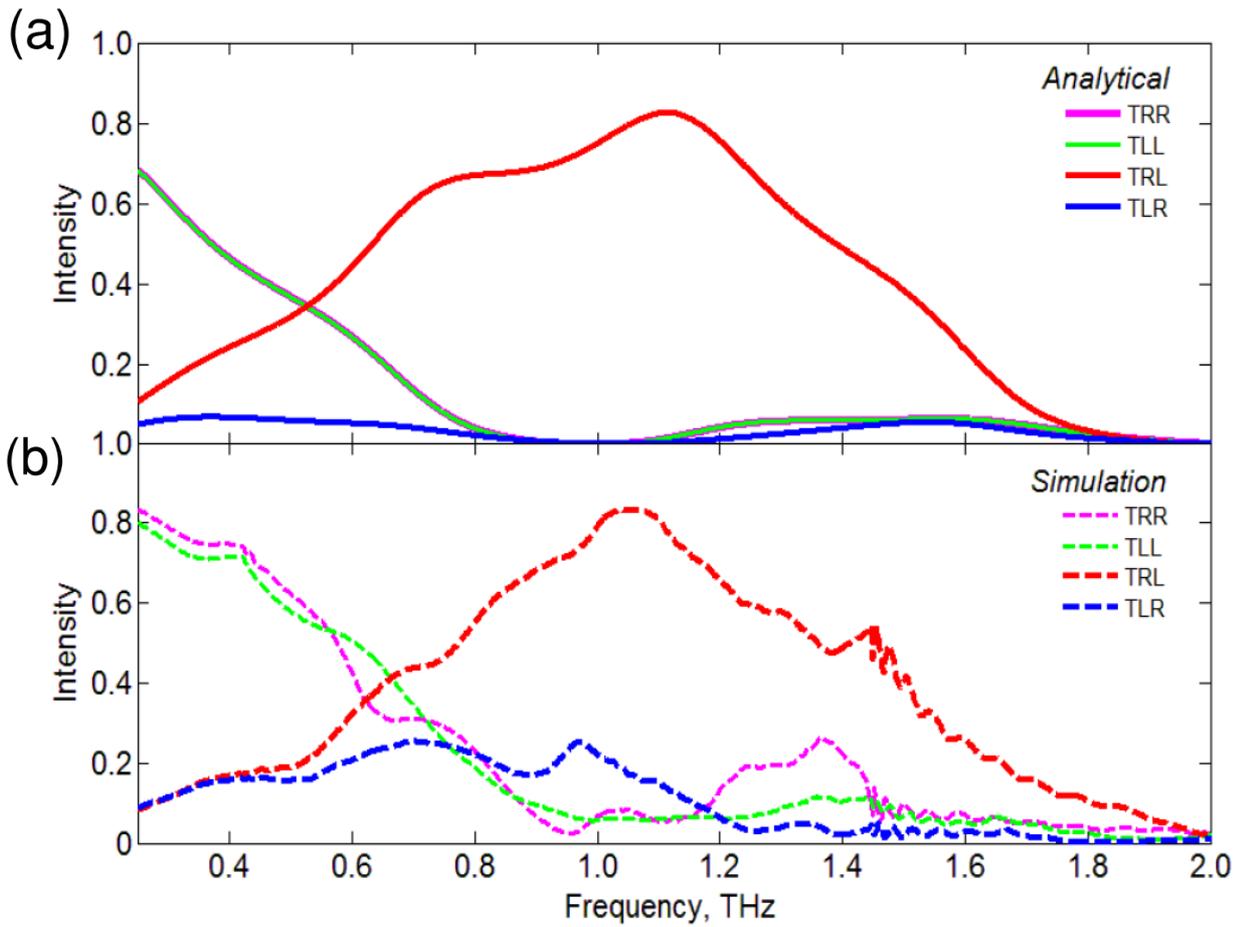

Figure 2. (a) The analytically derived transmission responses for the Jones matrix in a circular basis. (b) Simulated circular Jones matrix responses for the dimensions given in Figure 1c for the herringbone device.

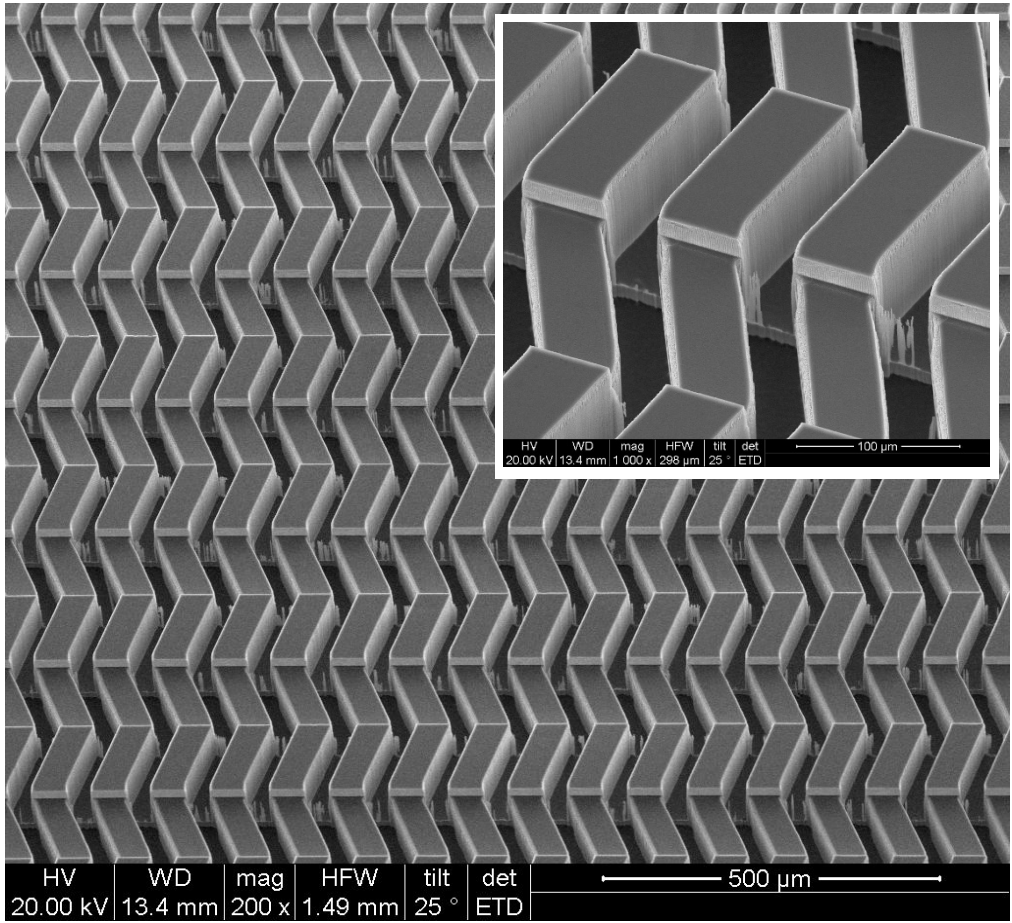

Figure 3: Scanning Electron Micrograph of the fabricated silicon herringbone device. Inset: Zoomed in Scanning Electron Micrograph showing the well-defined structures.

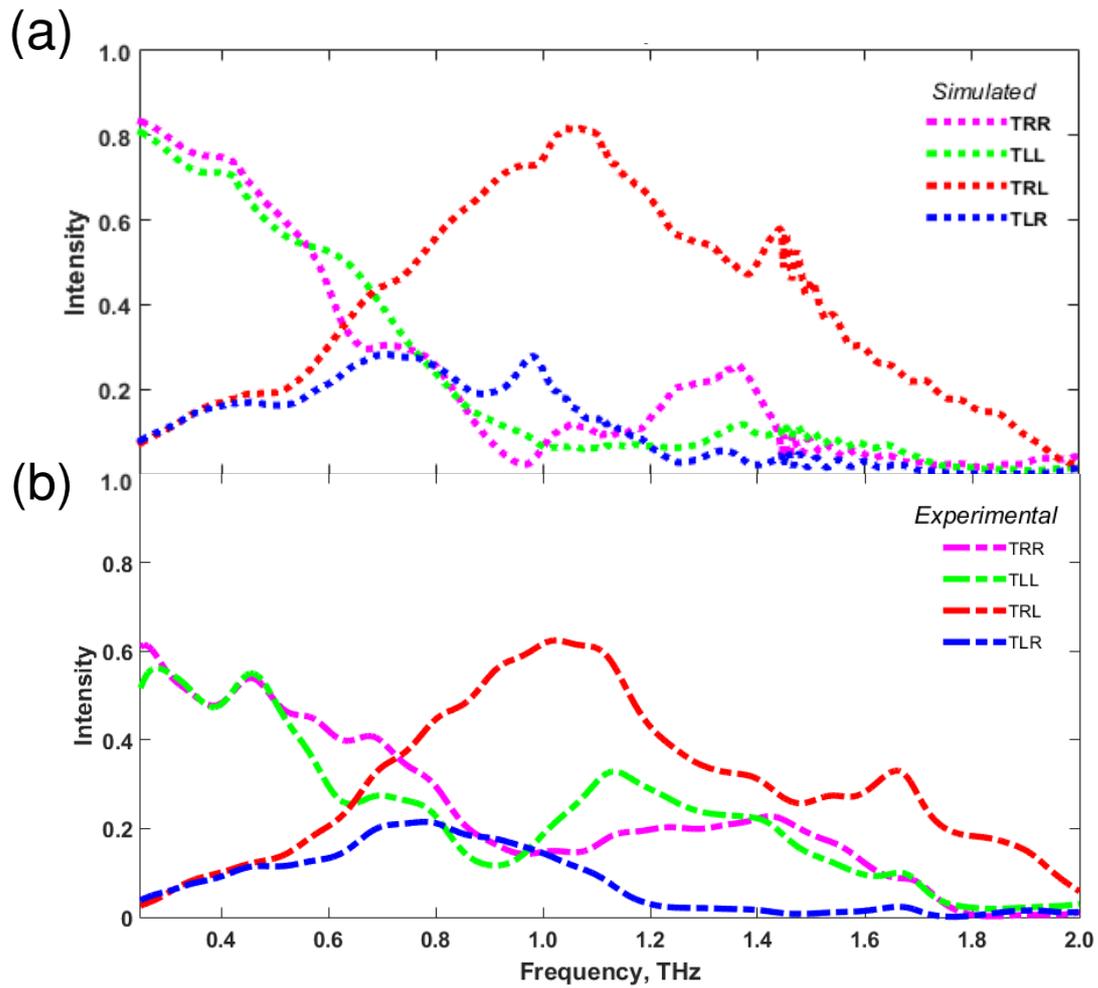

Figure 4. (a) Simulated responses of the device when using the dimensions of the fabricated device. (b) Experimentally obtained transmission coefficients from a THz-TDS system. The results are normalised to bare silicon.